\RequirePackage{fix-cm} 
\documentclass[a4paper,twoside,11pt,reqno]{amsart}

\usepackage{amsaddr,xcolor}
\usepackage{indentfirst,a4wide,verbatim}
\usepackage{amsmath,amsfonts,amssymb,amsgen,amsbsy,dsfont,eucal,mathrsfs,dsfont,stmaryrd}
\usepackage[square,numbers]{natbib}
\newcommand{\eqdef}{\stackrel{\text{\tiny{def}}}{=}} 

\newcommand{\ud}{\mathrm{d}}

\newcommand{\ui}{\mathrm{i}}
\newcommand{\um}{{\bar{u}}}

\newcommand{\half}{{\textstyle{1\over2}}}

\newcommand{\third}{{\textstyle{1\over3}}}

\newcommand{\fifth}{{\textstyle{1\over5}}}

\renewcommand{\Re}{\operatorname{Re}}

\title{Remarks on dispersion-improved shallow water equations with uneven bottom}  

\author[D. Clamond]{Didier Clamond}

\address{Universit\'e C\^ote d'Azur, CNRS UMR 7351, Laboratoire J. A. Dieudonn\'e,\\
Parc Valrose, F-06108 Nice cedex 2, France.}
\email{didier.clamond@univ-cotedazur.fr}

\begin{document}

\begin{abstract}
It is shown that asymptotically consistent modifications of (Boussinesq-like) shallow water 
approximations, in order to improve their dispersive properties, can fail for 
uneven bottoms (i.e., the dispersion is actually not improved). It is also shown that these 
modifications can lead to ill-posed equations when the water depth is not constant. These drawbacks are 
illustrated with the (fully nonlinear, weakly dispersive) Serre equations. We also derive 
asymptotically consistent, well-posed, modified Serre equations with improved dispersive 
properties for constant slopes of the bottom. 
\end{abstract}

\maketitle


\medskip


\section{Introduction}\label{intro}

Shallow water models are weakly dispersive approximations of the Euler equations with a free surface. 
For many applications, it is necessary to improve their dispersive and shoaling properties, for instance 
modifying the equations in an asymptotically consistent way. This possibility is investigated here for 
the \citet{Serre1953} equations with uneven bottom, this special case being of general interest and it 
is sufficient to illustrate the problem for any other shallow water approximation (e.g., Boussinesq-like 
equations \citep{Kirby2016,MadsenSchaffer1999}).

We then consider two-dimensional surface gravity waves in irrotational motion propagating 
at the surface of perfect incompressible fluid. The free surface at $y=\eta(x,t)$ and the 
bottom at $y=-d(x,t)$ are both impermeable, with $x$ the horizontal variable, $y$ 
the vertical upper one and $t$ the time; the still water level is at $y=0$. 
Assuming long waves in shallow water (i.e., $\partial_x$ and $\partial_t$ are ``small'' 
operators) without restriction on their amplitudes (i.e.,  fully nonlinear), \citet{Serre1953} 
derived a set of approximate equations for constant depth. In presence of a varying bottom, 
these equations can be written \citep{SSantos1985}
\begin{align}
h_t\ +\,\left[\,h\,\bar{u}\,\right]_x\ &=\ 0, \label{mass}\\
\um_t\ +\ \um\,\um_x\ +\ g\,\eta_x\ &=\ \half\left(\tilde{\gamma}+\breve{\gamma}\right)
d_x\ -\ \third\,h^{-1}\left[\,h^2\,\tilde{\gamma}\,+\,
\half\,h^2\,\breve{\gamma}\,\right]_x, \label{momnoncons}
\end{align}  
where $h=\eta+d$ is the total water depth, $\um$ is the depth-averaged horizontal 
velocity, $g>0$ is the acceleration due to gravity, $\breve{\gamma}$ and $\tilde{\gamma}$ 
being the (approximate) vertical accelerations at, respectively, the bottom and the free 
surface, i.e.,
\begin{align} 
\breve{\gamma}\, =\,  -d_{tt}\, -\, 2\,\um\,d_{xt}\, -\, \um^2\,d_{xx}\, -\, 
(\,\um_t\,+\,\um\,\um_x\,)\,d_x, \quad 
\tilde{\gamma}\, =\, \breve{\gamma}\, +\, h\left\{\,\um_x^{\,2}\, -\, \um_{xt}\, -\, \um\,\um_{xx}
\,\right\}.\label{defgammasur}
\end{align}
The equation (\ref{mass}) for the mass conservation is exact, while the momentum 
equation (\ref{momnoncons}) is an approximation: its left-hand side involves first-order 
terms (in shallowness parameter, i.e., in derivatives) and its right-hand side involves 
third-order terms. This approximation yields a fully nonlinear but only weakly dispersive 
model of water waves \citep{Wu2001}. 
It is then desirable to improve these models for a better description of dispersive 
effects without increasing the mathematical complexity, that is without introducing 
higher-order derivatives because they are computationally demanding and prone 
to trigger numerical instabilities. 

Thus, in order to improve the dispersive properties of shallow water models --- i.e., to extend 
their validity to deeper water --- some asymptotically consistent modifications of the 
momentum equation have been proposed. These modifications involve free parameters that 
can be chosen to tune the (weakly dispersive) linear dispersion relation such that it better 
matches the (fully dispersive) exact relation 
\citep{AntunesdoCarmo2013b,ClamondEtAl2017,MadsenEtAl1991,Nwogu1993,Witting1984}. 
For the Serre equations, such modified equations can be obtained replacing the momentum 
equation (\ref{momnoncons}) by  \citep{CienfuegosEtAl2006,MadsenEtAl1991} 
\begin{equation}\label{momnonconsmod}
\left(1-\alpha\/d^2\/\partial_x^{\,2}\right)\left(\/\um_t\/+\/\um\/\um_x\/ 
+\/g\/\eta_x\/\right) =\, \half\left(\tilde{\gamma}+\breve{\gamma}\right)
d_x\, -\, \third\/h^{-1}\/\partial_x\left(\/h^2\/\tilde{\gamma}\/+\/\half\/h^2\/\breve{\gamma}\/\right),
\end{equation}
where $\alpha$ is a free parameter at our disposal. One can easily check that $\alpha=0$ yields 
\eqref{momnoncons} and that equation (\ref{momnonconsmod}) with $\alpha\neq0$ is asymptotically 
consistent with  (\ref{momnoncons}). 
The parameter $\alpha$ is generally chosen considering a travelling wave of permanent 
form in constant depth. In such a case, the linear dispersion relation (relating the 
angular frequency $\omega$ and the wavenumber $k$) of the linearised Serre equations is a 
$(2,2)$-Pad\'e approximation (in the wavenumber) of the exact relation, while when $\alpha\neq0$ 
one gets the  $(4,2)$-Pad\'e approximation 
\begin{equation}\label{DRSM}
d\,g^{-1}\,\omega^2\ =\,\left[\,(kd)^2\,+\,\alpha\,(kd)^4\,\right]\left[\,1\,+\,(\third+\alpha)\,
(kd)^2\,\right]^{-1}.
\end{equation}
For all $\alpha$, the linear dispersion relation \eqref{DRSM} matches the exact one $\omega^2
=gk\tanh(kd)$ at least up to the fourth-order in its Maclaurin expansion in terms of the wavenumber, 
but for $\alpha=1/15$ the matching is up to the sixth-order.
Thus, $\alpha=1/15$ is the best choice to improve the dispersive properties (according 
to the criterion considered here). 

However, this improvement occurs only for horizontal bottoms. Indeed, in presence of a bottom 
slope ($d_x\neq0$), the dispersive properties of the modified equations are of fourth-order 
only for all $\alpha$. This means that the modification (\ref{momnonconsmod}) cannot 
improve the model in presence of a varying seabed, as one can check (see below). 
(In practice, the dispersive properties are nevertheless somewhat improved for 
very mild slopes ($|d_x|\ll1$), but it is of practical interest to remove this 
restriction.) Worse, we show below that \eqref{momnonconsmod} generally yields ill-posed 
models for varying bottoms.

In this note, we propose an alternative modification of the Serre (and others) momentum 
equation such that the dispersive effect are improved for finite constant slopes. To this 
aim, we  first reduce, in section \ref{appslope}, the exact (i.e., fully dispersive) linear 
equations to a single pseudo--differential equation for the free surface only. 
In section \ref{secmSerre}, we consistently modify the Serre equations with an 
unknown operator that is determined by identification with the shallow water approximation  
of the exact linear equation. These modified Serre equations for varying depth 
thus provide an improvement, at least for mild curvature of the seabed (i.e., 
when the gradient of the bottom slope is very small). In section \ref{welpos}, we show 
that these modified Serre equations are linearly well-posed for constant slopes, while the 
modified equations using \eqref{momnonconsmod} with $\alpha=1/15$ are ill-posed for varying bottoms.

\section{Linear waves on sloping beach}\label{appslope}

The (fully dispersive) linearised equations for an irrotational motion \citep{Stoker1957} are
\begin{align}
\phi_{xx}\ +\ \phi_{yy}\ &=\ 0 \qquad\text{for}\quad\! -d\leqslant y\leqslant0, \label{lineqstanFD1}\\
\phi_y\ +\ d_t\ +\ d_x\,\phi_x\ &=\ 0 \qquad\text{\,at}\quad y=-d, \\
\phi_t\ +\ g\,\eta\ &=\ 0 \qquad\text{\,at}\quad y=0, \label{bersurlin}\\
\phi_y\ -\ \eta_t\ &=\ 0 \qquad\text{\,at}\quad y=0, \label{lineqstanFD4}
\end{align}
where $\phi$ is a velocity potential. Here, we consider constant slopes, i.e., $d(x,t)=
d_0+sx$ where $d_0\geqslant0$ is the depth at $x=0$ and $s\eqdef\tan(\theta_0)\geqslant0$ 
is the constant slope of the seabed ($-\theta_0$ the seabed angle of inclination). 
Using the conformal mapping $z=x+\ui\/y\mapsto Z=X+\ui\/Y$ where 
\begin{equation}
Z\ \eqdef\ \frac{d_0}{\theta_0}\,\log\!\left(\frac{z}{d_0}+\frac{1}{s}\right),
\end{equation}
the wedge domain $\{sx\geqslant-d_0;-d_0-sx\leqslant y\leqslant0\}$ 
is mapped onto the strip $-d_0\leqslant Y\leqslant0$ with $x=-d_0/s\mapsto X=-\infty$ 
and $x=+\infty\mapsto X=+\infty$. 
In the mapped variables, after elimination of $\eta(x,t)=-g^{-1}\phi_t(x,0,t)$, the equations 
(\ref{lineqstanFD1})--(\ref{lineqstanFD4}) become
\begin{align}
\Phi_{XX}\ +\ \Phi_{YY}\ &=\ 0 \qquad\text{for}\quad\! -d_0\leqslant Y\leqslant0, \label{LapeqPhi}\\
\Phi_Y\ &=\ 0 \qquad\text{\,at}\quad Y=-d_0, \label{boteqPhi}\\
\Phi_Y\ +\ g^{-1}\,\theta_0\,\exp\!\left(\theta_0\/X/d_0\right)\Phi_{tt}\ &=\ 0 
\qquad\text{\,at}\quad Y=0,\label{sureqPhi}
\end{align}
where $\Phi(X,Y,t)\eqdef\phi(x(X,Y),y(X,Y),t)$ is the velocity potential in the mapped plane. 

The general solution of the Laplace equation (\ref{LapeqPhi}) satisfying the lower boundary 
condition (\ref{boteqPhi}) is (see \cite{Clamond1999,Clamond2003} for details) 
\begin{align}
\Phi(X,Y,t)\ &=\ \Re\!\left\{\breve{\Phi}(Z+\ui\/d_0,t)\right\}\ =\ 
\half\,\breve{\Phi}(Z+\ui\/d_0,t)\ +\ \half\,\breve{\Phi}(Z^*-\ui\/d_0,t) 
\nonumber\\ 
&=\ \cos\!\left((Y+d_0)\,\partial_X\right)\breve{\Phi}(X,t)\ =\ 
\sum_{n=0}^\infty\frac{(-1)^n}{(2n)!}\,(Y+d_0)^{2n}\,\frac{\partial^{2n}\,\breve{\Phi}(X,t)}
{\partial\/X^{2n}},\label{solPhibot}
\end{align}
where a star denotes the complex conjugate and $\breve{\Phi}(X,t)\eqdef\Phi(X,-d_0,t)$ is the 
velocity potential at the bottom.
Substituting (\ref{solPhibot}) into the upper boundary condition (\ref{sureqPhi}), one 
gets the complex difference-differential equation 
\begin{align}
\breve{\Phi}_X(X\!+\!\ui\/d_0,t)\, -\, \breve{\Phi}_X(X\!-\!\ui\/d_0,t)\, 
+\, \frac{\theta_0}{\ui\/g}\,\exp\!\left(\!\frac{\theta_0\/X}{d_0}\!\right)\!\left[\/
\breve{\Phi}_{tt}(X\!+\!\ui\/d_0,t)\/+\/\breve{\Phi}_{tt}(X\!-\!\ui\/d_0,t)\/\rule{0mm}{4mm}\right] 
=\/ 0.
\label{ddceq}
\end{align}
Exploiting the relation $\breve{\Phi}(X\pm\ui\/d_0,t)=\exp\!\left(\pm\/\ui\,d_0\,\partial_X\right) 
\breve{\Phi}(X,t)$ (Taylor expansion around $d_0=0$) and using the velocity potential at the 
free surface $\tilde{\Phi}(X,t)\eqdef\Phi(X,0,t)=\cos\!\left(d_0\/\partial_X\right) \breve{\Phi}(X,t)$, 
the difference-differential \eqref{ddceq} equation is rewritten as the real pseudo-differential 
equation
\begin{equation}\label{pseudoeq1sur}
\tan\!\left(d_0\,\partial_X\right)\tilde{\Phi}_X(X,t)\ -\ g^{-1}\,\theta_0\,\exp\!
\left(\theta_0\,X/d_0\right)\tilde{\Phi}_{tt}(X,t)\ =\ 0.
\end{equation} 
Note that $H(X,t)\eqdef \eta(x(X,0),t)=-g^{-1}\tilde{\Phi}_t(X,t)$ also satisfies 
the equation (\ref{pseudoeq1sur}) because the latter is linear and its coefficients 
are independent of $t$.

With the change of independent variable $\xi\eqdef d_0\exp(\theta_0 X/d_0)=x+d_0/s$ (hence 
$\ud\/\xi/\ud\/X=\theta_0\xi/d_0$) and since $\tilde{\Phi}(X,t)=\phi(\xi-d_0/s,0,t)$ is the 
velocity potential at the free surface, the equation (\ref{pseudoeq1sur}) becomes
\begin{equation}\label{DRexactxi}
\left\{\,\frac{\partial}{\partial\/\xi}\,\tan\!\left(\theta_0\,\xi\,\frac{\partial}
{\partial\/\xi}\right)-\,\frac{1}{g}\,\frac{\partial^2}{\partial\/t^2}\,\right\}
\phi(\xi-d_0/s,0,t)\ =\ 0,
\end{equation}
where 
\begin{align}
\tan\!\left(\!\theta_0\,\xi\,\frac{\partial}{\partial\/\xi}\!\right)\eqdef\, \theta_0\,
\xi\,\frac{\partial}{\ud\/\xi}\, +\, \frac{\theta_0^{\,3}}{3}\,\xi\,\frac{\partial}
{\partial\/\xi}\,\xi\,\frac{\partial}{\partial\/\xi}\,\xi\,\frac{\partial}{\partial\/\xi}
\, +\, \frac{2\,\theta_0^{\,5}}{15}\,\xi\,\frac{\partial}{\partial\/\xi}\,\xi\,\frac{\partial}
{\partial\/\xi}\,\xi\,\frac{\partial}{\partial\/\xi}\,\xi\,\frac{\partial}{\partial\/\xi}\,
\xi\,\frac{\partial}{\partial\/\xi}\ +\ \cdots.
\end{align}
Finally, the equation (\ref{DRexactxi}) differentiated with respect of $t$ is rewritten
--- since $\partial_\xi=\partial_x$, $d=s\xi=d_0+sx$, $s=\tan(\theta_0)=d_x$ and 
$\phi_t(x,0,t)=-g\eta(x,t)$ ---  in terms of the original variables 
\begin{equation}\label{DRexactx}
\left\{\,\frac{\partial^2}{\partial\/t^2}\,-\,g\,\frac{\partial}{\partial\/x}\,\tan\!\left(
\frac{\arctan(d_x)}{d_x}\,d\,\frac{\partial}{\partial\/x}\right)\right\}
\eta(x,t)\ =\ 0.
\end{equation}
It should be emphasised that the equation (\ref{DRexactx}) is exact for (fully dispersive) 
linear waves provided that $d_{xx}=d_t=0$, i.e., for flat static bottoms of arbitrary inclination, 
in particular for a constant depth. For more general seabeds, (\ref{DRexactx}) provides a 
reasonable (linear) approximation if the bottom varies very slowly in time and if its curvature 
is small, such assumptions being made to derive shallow water approximations. 

The relation  (\ref{DRexactx}) suggests to introduce an ``apparent'' or ``effective'' 
water depth $D$ as 
\begin{equation}
D\ \eqdef\ d_x^{\,-1}\,\arctan(d_x)\/\,d\ \leqslant\ d.
\end{equation}
This shows that a bottom slope creates a slowdown of the wave compared to a horizontal
bottom of the same depth, an effect conjectured in \cite{DutykhClamond2011} from 
a non-dispersive shallow water model. 
In shallow water $\partial_x$ is ``small'' and, since $d_{xx}=0$, the operator 
$\tan\!\left(\/D\,\partial_x\/\right)$ can be expanded up to the fifth-order as
\begin{align}
\tan\!\left(D\,\partial_x\right)\,&\eqdef\ D\,\partial_x\ +\ \third\,D\,\partial_x\,D\,
\partial_x\,D\,\partial_x\ +\ {\textstyle{\frac{2}{15}}}\,D\,\partial_x\,D\,\partial_x\,
D\,\partial_x\,D\,\partial_x\,D\,\partial_x\ +\ \cdots\quad \nonumber\\
&\approx\ d\,\partial_x\ +\ \third\,\partial_x\,d^3\,\partial_x^{\,2}\ +\ 
{\textstyle{\frac{2}{15}}}\,d^5\,\partial_x^{\,5}\ +\ {\textstyle{\frac{4}{3}}}\,d_x\,
d^4\,\partial_x^{\,4}\ +\ 3\,d_x^{\,2}\,d^3\,\partial_x^{\,3}\ +\ d_x^{\,3}\,d^2\,
\partial_x^{\,2}. \label{tanDsw}
\end{align}
This expansion should be compared with the (approximate) shallow water equations in order 
to derive suitable improvements. We note in passing that some approximate and empirical 
relations used in engineering for water waves propagating over mild slopes may be somewhat 
improved replacing $d$ by $D$. Similar modifications have already been proposed 
\citep{Ehrenmark2005,Porter2020,Porter2019} and computations therein suggest that 
the effective depth can improve significantly the performance of a model. 

It should be noted that the equation (\ref{DRexactx}) can be solved analytically for various 
wave fields over constant slope \cite{Ehrenmark2005,Roseau1958}. Then, in order to improve a 
shallow water model, similar solutions must be derived for this approximation. Such special 
analytic solutions, often complicated or even impossible to write in closed-form, are actually 
not needed for improving the linear dispersion. Indeed, it is sufficient, simpler 
and more general to compare the equations directly via their shallow water expansions, i.e., 
expansions such as (\ref{tanDsw}) obtained assuming that $\partial_x$ and $\partial_t$ are 
``small'' operators, as illustrated below.

\section{Modified Serre's equations}\label{secmSerre}

In order to address the drawback of (\ref{momnonconsmod}), a better modification is sought 
replacing $-\alpha\,d^2\,\partial_x^{\,2}$ by another  
second-order differential operator $\mathscr{D}$ to be defined such that the linearised 
equations match the (fully dispersive) exact ones for constant slopes up to the highest 
possible order in the shallowness expansion. Of course, this alternative modification of 
the momentum equation is also asymptotically consistent with the original equation, as one 
can easily check.

We thus consider infinitesimal waves with a fluid motion close to rest --- i.e., $\eta$ 
and $\um$ are small --- with $d=d(x)$ and $d_{xx}=0$. The linearised modified Serre-like 
equations are then 
\begin{gather}
\eta_t\ +\,\left[\,d\,\bar{u}\,\right]_x\ =\ 0, \label{maslinslopemodserre}\\
\left(1+\mathscr{D}\right)\left(\,\bar{u}_t\,+\, g\,\eta_x\,\right)\, -\ \third\,d^2\,\bar{u}_{xxt}\ 
-\ d_x\,d\,\bar{u}_{xt}\ = \ 0, \label{mommodserre}
\end{gather}
and eliminating $\bar{u}$ between these two relations, one gets after some algebra 
\begin{equation}\label{modSerLineta}
\left\{\,g^{-1}\,\partial_t^{\,2}\,-\,\partial_x\,d\left[\,d\,+\,d\,\mathscr{D}\,-\,
\third\,\partial_x\,d^3\,\partial_x\,\right]^{-1}\left(\,d\,+\,d\,\mathscr{D}\,\right)
\partial_x\,\right\}\eta\ =\ 0.
\end{equation}
The sixth-order shallow water expansion (i.e., assuming that $\partial_x$ is ``small'') 
of this linear equation
\begin{align}
\left\{\,g^{-1}\/\partial_t^{\,2}\,-\,\partial_x\/d\/\partial_x\,-\,\third\/\partial_x^{\,2}
\/d^3\/\partial_x^{\,2}\,+\,\third\/\partial_x\/d\/\mathscr{D}\/d^{-1}\/\partial_x\/d^3\/
\partial_x^{\,2}\, -\,{\textstyle{1\over9}}\/\partial_x^{\,2}\/d^3\/
\partial_x\/d^{-1}\/\partial_x\/d^3\/\partial_x^{\,2}\,\right\}\eta\, \approx\, 0, \label{DRSWmS}
\end{align}
is to be compared with the exact (i.e., fully dispersive) linear relations  
(\ref{DRexactx})--(\ref{tanDsw}) for constant slopes. Thus, the expansion 
(\ref{DRSWmS}) matches the exact one up to the sixth-order only if 
\begin{equation}\label{defDmS6th}
\mathscr{D}\ =\ {\textstyle{1\over5}}\,d_x^{\,2}\ -\ {\textstyle{1\over15}}\,d^{\/-1}
\,\partial_x\,d^3\,\partial_x. 
\end{equation}
The classical improvement is recovered on constant depth, as it should be. This choice 
is an improvement for constant slopes but, in practice, it should also improve the model 
when the bottom curvature is small (i.e., if $|d\/d_{xx}|\ll1$) and, at least, when the 
bottom varies very slowly in time.

The operator $\mathscr{D}$ derived here in the expression \eqref{defDmS6th} is, obviously, 
also suitable for any shallow water approximations having the same linear approximation  
as the Serre equations.

\section{Linear well-posedness}\label{welpos}

With \eqref{defDmS6th}, the differential operator $d+d\mathscr{D}$ is a self-adjoint and 
positive-definite, and the linear equation \eqref{modSerLineta} yields
\begin{equation}\label{modSerLinetasimp}
\left\{\,g^{-1}\,\partial_t^{\,2}\,-\,\partial_x\,d\left[\,d\,+\,\fifth\,d\,d_x^{\,2}
\,-\,{\textstyle{2\over5}}\,\partial_x\,d^3\,\partial_x\,
\right]^{-1}\left(\,d\,+\,d\,\mathscr{D}\,\right)\partial_x\,\right\}\eta\ =\ 0.
\end{equation} 
The differential operator inside the square bracket of \eqref{modSerLinetasimp} is 
also self-adjoint and positive-definite (since $d>0$), so it is invertible and  
the equation \eqref{modSerLinetasimp} is therefore well defined. 

Conversely, with the operator $\mathscr{D}_0\eqdef-{\textstyle{1\over15}}\/
d^2\/\partial_x^{\,2}$ commonly used to improve the dispersion, $d+d\mathscr{D}_0$ is not 
self-adjoint and not positive-definite if $d_x\neq0$, and the equation \eqref{modSerLineta} 
becomes 
\begin{equation}\label{modSerLinetawrong}
\left\{\,g^{-1}\,\partial_t^{\,2}\,-\,\partial_x\,d\left[\,d\,+\,\fifth\,d^2\,d_x\,\partial_x
\,-\,{\textstyle{2\over5}}\,\partial_x\,d^3\,\partial_x\,\right]^{-1}
\left(\,d\,+\,d\,\mathscr{D}_0\,\right)\partial_x\,\right\}\eta\ =\ 0.
\end{equation} 
The differential operator inside the square bracket of \eqref{modSerLinetawrong} is 
not positive-definite (nor negative-definite) if $d_x\neq0$, so it is not 
invertible and the equation \eqref{modSerLinetawrong} is ill-posed, in general. 
Thus, depending to the problem at hand, unphysical phenomena and numerical instabilities 
may appear as a consequence of ``improving'' the shallow water equations using $\mathscr{D}_0$.  

This simple analysis shows that the modification \eqref{momnonconsmod} is generally leading 
to ill-posed models for uneven bottoms, that is a worse drawback than not being a dispersion 
improvement.

\section{Discussion}\label{secconclu}

We have shown that attempts to improve shallow water models, by tweaking the equations in an 
asymptotically consistent way, does not always lead to sound improved models. Worse, one can 
get ill-posed equations. These points were illustrated with the Serre equations where their 
classical modification \eqref{momnonconsmod} improves the dispersion and it is well-posed only 
in contant depth; for varying bottoms, \eqref{momnonconsmod} does not improve the dispersion 
and it is generally linearly ill-posed. 
 
Considering fully-dispersive linear waves on constant slopes, we proposed another modification 
of the Serre equations such that their dispersive properties are improved for constant slopes. 
We also showed that this modification is linearly well-posed. 

Here, we focused on Serre's equations but similar modifications hold for any variant of shallow 
water (Boussinesq-like) approximations. The way the equations are modified to consistently improve 
the dispersion is of course not unique. Whatever approach is considered, suitable modifications 
can be derived along the lines illustrated here for the Serre equations. 

The effectiveness, for practical applications, of this approach could be investigated via numerical 
simulations. These are not the purpose of the present short note where we focussed on demonstrating 
theoretically the drawbacks of \eqref{momnonconsmod} and how they can be addressed.  Further 
theoretical investigations could be performed. For instance, one can look for:  (i) dispersion 
improvements for arbitrary (i.e., not only constant) slopes; (ii) a three-dimensional extension 
(i.e., two horizontal spacial dimensions); (iii) variational derivations of modified equations.  
These will be the purpose of future investigations.


\end{document}